\documentclass[twocolumn,authoryear]{autart}    

\usepackage{graphicx}          

\usepackage{mathptmx} 

\usepackage{amsmath} 

\usepackage{amssymb}  

\usepackage{graphicx}
\usepackage{ntheorem}
\usepackage{color}
\usepackage{dsfont}
\usepackage[authoryear,round]{natbib}

\newtheorem{thm1}{Theorem}

\newtheorem{cor1}{Corollary}

\newtheorem{lem1}{Lemma}

\newtheorem{defn1}[thm]{Definition}

\newtheorem{rem1}[thm]{Remark}

\newcommand{\bsmtx}{\left[ \begin{smallmatrix}} 
    \newcommand{\esmtx}{\end{smallmatrix} \right]} 
\newcommand{\bmtx}{\begin{bmatrix}}
  \newcommand{\emtx}{\end{bmatrix}}

\begin{document}

\begin{frontmatter}
\title{Conditions for the equivalence between IQC and graph separation stability results}\author[JC]{Joaquin Carrasco}\ead{joaquin.carrascogomez@manchester.ac.uk}\ and
\author[PS]{Peter Seiler}\ead{seile017@umn.edu}
\address[JC]{Control Systems Centre, School of Electrical and Electronic Engineering,
	The University of Manchester,
	Manchester M13 9PL, UK}
\address[PS]{Aerospace Engineering and Mechanics Department, University of Minnesota, 107 Akerman Hall, 110 Union St. SE Minneapolis, MN 55455-0153}

\begin{abstract}                          

  This paper provides a link between time-domain
  and frequency-domain stability results in the literature.
  Specifically, we focus on the comparison between stability results for a feedback interconnection of two nonlinear systems stated in terms of frequency-domain conditions. While the Integral Quadratic Constrain (IQC) theorem can cope with them via a homotopy argument for the Lurye problem, graph separation results require the transformation of the frequency-domain conditions into truncated time-domain conditions. 
  To date, much of the literature
  focuses on ``hard'' factorizations of the multiplier, considering
  only one of the two frequency-domain conditions. Here it is shown
  that a symmetric, ``doubly-hard'' factorization is required to
  convert both frequency-domain conditions into truncated time-domain conditions. By using the appropriate factorization, a novel comparison between the results obtained by IQC and separation theories is then provided. As a result, we identify under what conditions the IQC theorem may provide some advantage.
\end{abstract}

\end{frontmatter}

\section{Motivation}

Classical multiplier theory is a well known technique to reduce the
conservatism of absolute stability
criteria~\citep{Zames68,desoer75}. Frequency-domain and time-domain conditions are combined, and the canonical factorization of the
multiplier is the essential tool to ensure that time-domain properties
can be recovered from the frequency-domain
conditions~\citep{jonssonphd,Goh95,goh96b,Carrasco12a}. 

The IQC theorem by~\cite{Megretski97} uses only frequency-domain inequalities and provides a shortcut to avoid conditions on the existence of factorizations by using a homotopy argument in their proof. However the original IQC framework was developed using
time-domain constraints by~\cite{yakubovich65,yakubovich67,yakubovich71}, so \cite{Megretski97} have coined the terms \emph{soft} and \emph{hard} IQC\footnote{It has been shown in \citep{Seiler:10,Seiler:15} that the same IQC can
	be either hard or soft depending on the factorization used to convert
	from frequency to time-domain; therefore the terms hard and soft factorizations terminology must be introduced.} 
to establish the connection between their IQC theorem and Yakubovich's work. Loosely speaking:
\begin{itemize}
\item an IQC is \emph{hard} when the time-domain version of the
  constraint holds for any finite time interval $[0,T]$;
\item an IQC is \emph{soft} when the time-domain version of the
  constraint holds for the interval $[0,\infty)$ but need not
  be satisfied on finite time intervals.
\end{itemize}

It may appear that a hard
factorization is equivalent to the canonical factorization in the
classical multiplier theory. In other words, one may think that a hard
factorization of an IQC is sufficient to convert frequency-domain
stability conditions to equivalent time-domain
conditions~\citep{goh96b,Seiler:10}. However, it has been shown that hard factorizations are not enough to establish such equivalence~\citep{Veenman:13b,Seiler:15}. The equivalence between IQC and the so-called dissipative inequality is shown in~\cite{Seiler:15}. The term hard factorization is still used, and then an extra condition is imposed on the solution of an LMI involving the LTI system.

The graph separation framework~\citep{safonov80,Teel:96,Georgiou:97} can be seen as a generalisation of the classical multiplier theory and uses truncated time-domain conditions to obtain stability result. Recently, \cite{Carrasco:15}  have shown that it is possible to establish a counterpart of the IQC theorem using the graph separation framework. However, they rely on results in~\citep{Seiler:15} and require one of the two systems in the interconnection to be LTI. 

This paper builds on the results presented in~\citep{Seiler:15} and~\cite{Carrasco:15}. The main contribution of this paper is the development of the counterpart of Lemma 2 in~\citep{Seiler:15}. With this new result, we can establish the equivalence between frequency-domain conditions and truncated time-domain conditions from a pure input-output point of view, without involving LMIs; hence the definition of the factorization does not require one of the systems to be LTI. This approach provides new insights, in particular, we are able to establish a formal comparison between stability results using IQC and graph separation theories for the feedback interconnection of two nonlinear systems.

The structure of the paper is as follows. Sections 2 and 3  provides the IQC theorem and discusses classical hard and soft factorizations as defined  by~\cite{Megretski97}. Section 4 states a new factorization, the so-called doubly-hard factorization, and characterises this factorization for a class of multipliers. Section 5 demonstrates that not all hard factorizations are doubly-hard factorizations. Section 6 develops two results for the stability of the feedback interconnection of two systems, one using the IQC theorem, and another using the graph separation result by~\cite{Teel:96}. Finally, Section 7 gives the conclusions of the paper. We use the same notation as in~\citep{Megretski97}.

\section{IQC theorem}
Definitions and results related with the IQC framework are given in this section.
\begin{defn1}	
A stable and causal system
$\Delta: \mathcal{L}_{2e}^m[0,\infty) \rightarrow \mathcal{L}_{2e}^l[0,\infty)$ is said to
satisfy the IQC defined by a bounded, measurable Hermitian-valued function $\Pi:j\mathds{R}\rightarrow\mathds{C}^{(m+l)\times(m+l)}$ if
\begin{equation}
\label{eq:IQCfreq}
\int_{-\infty}^{\infty}
\begin{bmatrix}\widehat{u}(j\omega)\\ \widehat{\Delta u}(j\omega)\end{bmatrix}^*
\Pi(j\omega)
\begin{bmatrix}\widehat{u}(j\omega)\\ \widehat{\Delta u}(j\omega)\end{bmatrix}
d\omega\geq 0,
\end{equation}	
for any $u\in\mathcal{L}_2^m[0,\infty)$. 
\end{defn1}
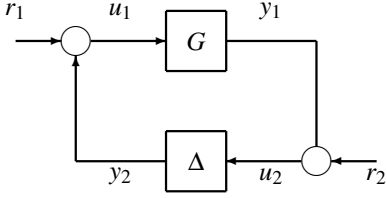
\begin{figure}[t]
	\centering
	\ifx\JPicScale\undefined\def\JPicScale{0.8}\fi
\unitlength \JPicScale mm
\begin{picture}(60,30)(0,0)
\linethickness{0.3mm}
\put(0,25){\line(1,0){7.5}}
\put(7.5,25){\vector(1,0){0.12}}
\linethickness{0.3mm}
\put(10,5){\line(1,0){15}}
\linethickness{0.3mm}
\put(10,5){\line(0,1){17.5}}
\put(10,22.5){\vector(0,1){0.12}}
\linethickness{0.3mm}
\put(12.5,25){\line(1,0){12.5}}
\put(25,25){\vector(1,0){0.12}}
\linethickness{0.3mm}
\put(35,25){\line(1,0){15}}
\linethickness{0.3mm}
\put(50,7.5){\line(0,1){17.5}}
\linethickness{0.3mm}
\put(35,5){\line(1,0){12.5}}
\put(35,5){\vector(-1,0){0.12}}
\put(0,30){\makebox(0,0)[cc]{$r_1$}}

\put(42.5,30){\makebox(0,0)[cc]{$y_1$}}

\linethickness{0.3mm}
\put(25,30){\line(1,0){10}}
\put(25,20){\line(0,1){10}}
\put(35,20){\line(0,1){10}}
\put(25,20){\line(1,0){10}}
\linethickness{0.3mm}
\put(25,10){\line(1,0){10}}
\put(25,0){\line(0,1){10}}
\put(35,0){\line(0,1){10}}
\put(25,0){\line(1,0){10}}
\put(30,5){\makebox(0,0)[cc]{$\Delta$}}

\put(30,25){\makebox(0,0)[cc]{$G$}}

\linethickness{0.3mm}
\put(10,25){\circle{5}}


\put(17.5,30){\makebox(0,0)[cc]{$u_1$}}

\linethickness{0.3mm}
\put(50,5){\circle{5}}

\linethickness{0.3mm}
\put(52.5,5){\line(1,0){7.5}}
\put(52.5,5){\vector(-1,0){0.12}}
\put(60,2.5){\makebox(0,0)[cc]{$r_2$}}

\put(42.5,2.5){\makebox(0,0)[cc]{$u_2$}}

\put(17.5,2.5){\makebox(0,0)[cc]{$y_2$}}

\end{picture}
	\caption{Lurye problem}\label{mp1}
\end{figure}

\begin{thm1}[IQC theorem \citep{Megretski97}] 
  \label{thm:IQC}
  Let $G\in\mathbf{RH}^{m\times l}_\infty$, let
  $\Delta:\mathcal{L}^m_{2e}[0,\infty)\rightarrow\mathcal{L}^l_{2e}[0,\infty)$
  be a bounded causal operator, and let
  $\Pi:j\mathds{R}\rightarrow\mathds{C}^{(m+l)\times(m+l)}$ be a bounded,
  measurable Hermitian-valued function. Assume that:
  \begin{enumerate}
  \item for every $\tau\in[0,1]$, the interconnection of $G$ and $\tau\Delta$ is well-posed;
  \item for every $\tau\in[0,1]$, the IQC defined by $\Pi$ is satisfied by $\tau\Delta$;
  \item there exists $\epsilon>0$ such that
    \begin{equation}
      \label{eq:GPiG}
      \begin{bmatrix}
        G(j\omega)\\I 
      \end{bmatrix}^*\Pi(j\omega)\begin{bmatrix}
        G(j\omega)\\I 
      \end{bmatrix} < -\epsilon I\qquad \forall\omega\in\mathds{R}.
    \end{equation}		
  \end{enumerate}
  Then, the feedback interconnection of $G$ and $\Delta$ in
  Fig.~\ref{mp1} is stable.
\end{thm1}

The multiplier $\Pi$ is normally defined as a block 2-by-2 matrix,
i.e.
\begin{equation}
\Pi=\begin{bmatrix}
\Pi_{11} & \Pi_{12}\\
\Pi_{21} & \Pi_{22}
\end{bmatrix}.
\end{equation}
where $\Pi_{11}$ is $m\times m$ and $\Pi_{22}$ is $l \times l$.  Then
$\Pi(j\omega)$ is called a \emph{positive-negative multiplier} if
there exists $\epsilon>0$ such that $\Pi_{11}(j\omega) \geq \epsilon I_m$
and $\Pi_{22}(j\omega)\leq-\epsilon I_l$ $\forall \omega\in\mathds{R}$. 

In this note we restrict our attention to positive-negative rational multipliers
$\Pi\in\mathbf{RL}_\infty^{(m+l)\times (m+l)}$.

\section{Hard and soft factorizations}

The IQC in Equation~\ref{eq:IQCfreq} can be expressed in the time-domain
and  this leads to a characterization of the IQC as soft
or hard.  Specifically, let
$\Pi(j\omega)=\Psi^\top (-j\omega) M \Psi(j\omega)$ where $\Psi$ is a
causal and stable transfer function.  Such factorizations are not
unique but can be computed with state-space methods \citep{scherer00}.
With some abuse of notation we will use the same notation for the
transfer function and its corresponding stable operator. The
IQC-factorization $(\Psi,M)$ is said to be \emph{soft} if 
\begin{equation}
\label{eq:soft}
\int_{0}^{\infty}
\left(\Psi\begin{bmatrix}u\\ \Delta u\end{bmatrix}\right)^\top 
M 
\left(\Psi\begin{bmatrix}u\\ \Delta u\end{bmatrix} \right) 
dt\geq 0,
\end{equation}	
for any $u\in\mathcal{L}_2^m[0,\infty)$.\footnote{Note that the dependence of
	time has been suppressed in Equation~\ref{eq:soft} for simplicity.
	More precisely, this soft IQC is 
	$\int_0^\infty y^\top (t) M y(t) dt \ge 0$ where
	$y:=\Psi \bsmtx u \\ \Delta u \esmtx$.  The time dependence will
	similarly be dropped in other time-domain IQCs.}  The
frequency-domain constraint of Inequality~\ref{eq:IQCfreq} implies the
time-domain soft constraint of Inequality~\ref{eq:soft} by Parseval's
theorem. The factorization is said to be \emph{hard} if
\begin{equation}
\int_{0}^{T}\left(\Psi\begin{bmatrix}u\\ \Delta u\end{bmatrix}\right)^\top 
M \left(\Psi\begin{bmatrix}u\\ \Delta u\end{bmatrix} \right) dt\geq 0,
\end{equation}	
for any $u\in\mathcal{L}_{2e}^m[0,\infty)$ and any $T>0$. This condition for a
hard factorization is more restrictive. Specifically, all
factorizations of $\Pi$ are soft but only certain factorizations are
hard. It is now clear that the factorization step, i.e.
$\Pi=\Psi^\sim M \Psi$, is a key point as the same $\Pi$ can have
hard and soft factorizations. These are called $(\Psi,M)$-hard and
$(\Psi,M)$-soft factorizations~\citep{Seiler:10,Seiler:15}.  The terms
\emph{complete} and \emph{conditional} IQCs by \cite{Megretski10} are
generalizations of hard and soft IQCs. The hard/soft terminology will
be used here.

There are simple sufficient conditions for the existence of a hard
factorization~\citep{goh96b}.  For positive-negative multipliers, it is
always possible to find a hard factorization $(\Psi,J_{m,l})$:
\begin{equation}
\Psi=\begin{bmatrix}
\Psi_{11} & 0\\
\Psi_{21} & \Psi_{22}
\end{bmatrix} 
\text{ and }  
J_{m,l}=\begin{bmatrix}
I_{m} & 0\\
0      & -I_{l}
\end{bmatrix}
\end{equation}
where $\Psi_{11}, \Psi_{11}^{-1}, \Psi_{22}$ are stable rational
transfer functions. This ensures that the truncation of the IQC will
preserve its sign~\citep{goh96b}. This fact is shown via a simple
argument as
\begin{multline}
\label{eq:InfIQC}
\int_{0}^{\infty}
\begin{bmatrix}\Psi_{11}u\\ (\Psi_{21}+\Psi_{22}\Delta) u\end{bmatrix}^\top 
J_{m,l}
\begin{bmatrix}\Psi_{11}u\\ (\Psi_{21}+\Psi_{22}\Delta) u\end{bmatrix}dt
=\\ \|\Psi_{11}u\|^2-\|(\Psi_{21}+\Psi_{22}\Delta) u\|^2\geq 0.
\end{multline}	
Any input $u$ can be truncated on $[0,T]$ and extended over the
positive real line by selecting an artificial input $z([T,\infty))$
such that $\Psi_{11}\tilde{u}(t)=0$ for all $t>T$ where the piecewise input
$\tilde{u}$ is defined by
\begin{equation}
\tilde{u}(t)=\begin{cases}
u(t) &\text{ if } t\leq T,\\
z(t) &\text{ if } t>T.\\
\end{cases}
\end{equation}
The pair $(\tilde{u},\Delta \tilde{u})$ satisfies the infinite horizon
constraint of Inequality~\ref{eq:InfIQC}. Hence, by construction, the
pair $(u, \Delta u)$ satisfies the constraint over the finite horizon
$[0, T]$.  The key point in this construction is the stability of
$\Psi_{11}^{-1}$ since it ensures that the artificial
input $\tilde{u}$ belongs to $\mathcal{L}_2$ and
$\|\Psi_{11}u\|_T=\|\Psi_{11}\tilde{u}\|_T$.

It may initially appear that this truncation is sufficient to complete
a dissipativity (or graph separation) proof for stability.  However,
the role of the second IQC condition seems underappreciated in the
literature. Specifically Equation~\ref{eq:GPiG} in the IQC theorem
is equivalent to the following second time-domain IQC condition (because
both $G$ and $\Psi$ are LTI):
\begin{equation}
\int_{0}^{\infty}\left(\Psi\begin{bmatrix}Gu\\ u\end{bmatrix}\right)^\top 
M \left(\Psi\begin{bmatrix}Gu\\ u\end{bmatrix}\right) dt
< -\epsilon \|u\|^2.
\end{equation}	
All operators in this IQC condition are stable LTI systems and hence
the condition can be checked via an equivalent frequency-domain
condition. However, this does not imply that the sign of this
inequality will be preserved under finite-horizon truncations in the
time-domain.  In particular, the key difficulty is observed if we use
the triangular factorization along with the truncation arguments
introduced above:
\begin{multline}
-\int_{0}^{\infty}
\begin{bmatrix}\Psi_{11}Gu\\ (\Psi_{21}G+\Psi_{22}) u\end{bmatrix}^\top 
J_{m,l}
\begin{bmatrix}\Psi_{11}Gu\\ (\Psi_{21}G+\Psi_{22}) u\end{bmatrix} dt
=\\ 
\| (\Psi_{21}G+\Psi_{22}) u\|^2 -   \|\Psi_{11}Gu\|^2 > \epsilon \|u\|^2.
\end{multline}	
We now see the difficulties in creating an extension of the input once a
truncation $u([0,T])$ has been selected. The extension of the
piecewise input on $[T,\infty)$ must cancel
$(\Psi_{21}G+\Psi_{22}) \tilde{u}$ for any time after the
truncation. It may be possible in some cases, but in general this
leads to piecewise $\tilde{u}\not\in\mathcal{L}_2[0,\infty)$ since
$\Psi_{22}^{-1}$ is not stable. This problem is linked to the well
known difficulties of applying feedback linearisation to non-minimum
phase systems~\citep{isidori:13}.

\section{Doubly-hard IQC factorization}

It is possible to show that positive-negative multipliers have a more
useful factorization for the purposes of stability analysis. It is
shown by \cite{Seiler:15} that J-spectral factorizations can be
constructed for positive-negative multipliers,
i.e. $\Pi(j\omega)=\Psi^\top (-j\omega) J_{m,l} \Psi(j\omega)$ where
$\Psi$ and $\Psi^{-1}$ are both stable transfer functions. Moreover,
this factorization allows us to ensure that the signs of both IQCs are
preserved under truncation.

To the best of our knowledge this duality property of the
factorization has been overlooked. The argument by \cite{Seiler:15},
where the $J$-spectral factorization is given, was based on storage
function and dissipativity arguments. The factorization there was
still referred to as a hard factorization with a focus on the IQC
condition for $\Delta$, but the second condition was established in terms of the resulting LMI. Here we propose a more symmetric and
convenient definition where we do not require the construction of the LMI, so we are able to establish the properties of the factorization without invoking the linearity of one of the systems.

In the graph framework, it is standard to use the graph and the inverse graph~\citep{safonov80,Teel:96}. The standard IQC notation uses the graph of the system $\Delta$. To develop a symmetric formulation, we define the IQC over the inverse graph, henceforward ``inverse-graph IQC'' as follows:

\begin{defn1}[Inverse-graph IQC]	
A stable and causal system
$\Delta: \mathcal{L}_{2e}^l[0,\infty)\rightarrow \mathcal{L}_{2e}^m[0,\infty)$ is said to strictly
satisfy the \emph{inverse-graph IQC} defined by a bounded, measurable Hermitian-valued function $\Pi:j\mathds{R}\rightarrow\mathds{C}^{(m+l)\times(m+l)}$ if there exists $\epsilon>0$ such that
\begin{equation}
\label{eq:IQCfreq1}
\int_{-\infty}^{\infty}
\begin{bmatrix}\widehat{\Delta u}(j\omega)\\ \widehat{u}(j\omega)\end{bmatrix}^*
\Pi(j\omega)
\begin{bmatrix}\widehat{\Delta u}(j\omega)\\ \widehat{u}(j\omega)\end{bmatrix}
d\omega\leq -\epsilon\|u\|,
\end{equation}	
for any $u\in\mathcal{L}_2^l[0,\infty)$.
\end{defn1}

 If $\Delta$ is linear, then \eqref{eq:IQCfreq1} is equivalent to~\eqref{eq:GPiG} by using $\Delta$ instead of $G$.
 
Then we can state the definition of the factorization which will lead to an equivalence between frequency-domain conditions and truncated time-domain conditions: 
\begin{defn1}[Doubly-hard factorization]
For a given $\Pi:j\mathds{R}\rightarrow\mathds{C}^{(m+l)\times(m+l)}$, a factorization $(\Psi,M)$ is said to be a
\emph{doubly-hard IQC factorization} if the following two conditions
hold:
\begin{enumerate}
	\item for any bounded and causal $\Delta_1:\mathcal{L}^m_{2e}[0,\infty)\rightarrow\mathcal{L}^l_{2e}[0,\infty)$, the IQC condition
	\begin{equation}
	\int_{-\infty}^{\infty}
	\begin{bmatrix}\widehat{u}(j\omega)\\ \widehat{\Delta_1 u}(j\omega)\end{bmatrix}^*
	\Pi(j\omega)
	\begin{bmatrix}\widehat{u}(j\omega)\\ \widehat{\Delta_1 u}(j\omega)\end{bmatrix}
	d\omega\geq 0,
	\end{equation}	
	for all $u\in\mathcal{L}_2^m[0,\infty)$ implies that 
	\begin{equation}
	\int_{0}^{T}
	\left(\Psi\begin{bmatrix}u\\ \Delta_1 u \end{bmatrix}\right)^\top 
	M \left( \Psi\begin{bmatrix}u\\ \Delta_1 u\end{bmatrix}\right)
	dt\geq 0.
	\end{equation}	
	for any $u\in\mathcal{L}_{2e}^m[0,\infty)$ and any $T>0$, and
	\item for any bounded and causal $\Delta_2:\mathcal{L}^l_{2e}[0,\infty)\rightarrow\mathcal{L}^m_{2e}[0,\infty)$, the \emph{inverse-graph IQC} condition
	\begin{equation}\label{eq:sIQC}
		\int_{-\infty}^{\infty}
		\begin{bmatrix}\widehat{\Delta_2 u}(j\omega)\\ \widehat{ u}(j\omega)\end{bmatrix}^*
		\Pi(j\omega)
		\begin{bmatrix}\widehat{\Delta_2 u}(j\omega)\\ \widehat{ u}(j\omega)\end{bmatrix}
		d\omega\leq -\epsilon \|u\|^2,
	\end{equation}	
	for all $u\in\mathcal{L}_2^l[0,\infty)$ implies that
	\begin{equation}
	\int_{0}^{T}
	\left(\Psi\begin{bmatrix} \Delta_2u\\ u \end{bmatrix}\right)^\top 
	M 
	\left(\Psi\begin{bmatrix} \Delta_2u\\ u\end{bmatrix} \right)
	dt \leq -\epsilon \|u\|_T^2,
	\end{equation}	
	for any $u\in\mathcal{L}_{2e}^l[0,\infty)$ and any $T>0$.
\end{enumerate}
\end{defn1}
Finally, we show that the key property to obtain a doubly-hard factorization is the stability of both $\Psi$ and $\Psi^{-1}$. This result requires Lemma 2 in~\cite{Seiler:15}, and the development of a result for the inverse-graph condition~\eqref{eq:sIQC}. 

Let 
\begin{equation}
\Psi\sim\left[\begin{array}{c|cc}
A & B_v & B_w\\ \hline
C & D_v & D_w
\end{array}\right].
\end{equation}
Define the functional $J$ on $v\in\mathcal{L}_2[0,\infty)$, $w\in\mathcal{L}_2[0,\infty)$ and $x_0\in\mathds{R}^n$ as
\begin{equation}
J(v,w,x_0)=\int_{0}^{\infty} z(t)^\top Mz(t)dt
\end{equation}
subject to 
\begin{eqnarray*}
\dot{x}(t)&=&Ax(t)+B_v v(t)+B_w w(t), \quad x(0)=x_0;\\
z(t)&=&Cx(t)+D_v v(t)+D_w w(t).
\end{eqnarray*}
Define the upper value $\bar{J}(x_0)$ as
$$\bar{J}(x_0):=\inf_{v\in\mathcal{L}_2[0,\infty)}\sup_{w\in\mathcal{L}_2[0,\infty)} 
J(v,w,x_0),$$
and the lower value $\underbar{J}(x_0)$ as
$$\underbar{J}(x_0):=\sup_{w\in\mathcal{L}_2[0,\infty)}\inf_{v\in\mathcal{L}_2[0,\infty)} 
J(v,w,x_0).$$
\begin{lem1}[\cite{Seiler:15}]\label{lem_ori}
	Let $\Pi$ be a multiplier and $(\Psi,M)$ any factorization with $\Psi$ stable. Assume $\Delta_1$ is a causal bounded operator such that
	 	\begin{equation}
	 	\int_{-\infty}^{\infty}
	 	\begin{bmatrix}\widehat{v}(j\omega)\\ \widehat{w}(j\omega)\end{bmatrix}^*
	 	\Pi(j\omega)
	 	\begin{bmatrix}\widehat{v}(j\omega)\\ \widehat{w}(j\omega)\end{bmatrix}
	 	d\omega\geq 0, 
	 	\end{equation}
                for any $v\in\mathcal{L}_2[0,\infty)$ and $w=\Delta_1
                v$. Then for all $T\geq0$, for all
                $v\in\mathcal{L}_2[0,\infty)$, and $w=\Delta_1 v$, the signal
                defined by\linebreak[4]
                $z=\Psi(j\omega)\begin{bmatrix}\widehat{v}(j\omega)\\
                  \widehat{w}(j\omega)\end{bmatrix}$ satisfies
	 		\begin{equation}
	 		\int_{0}^{T}
	 		z(t)^\top 
	 		M z(t)
	 		dt\geq -\bar{J}(x(T)),
	 		\end{equation}
	 		where $x(T)$ denotes the state of the system $\Psi$ at the instant $T$ when driven by the inputs $(v,w)$ with null initial conditions. 
\end{lem1}
\vspace{5mm}
\begin{lem1}\label{lem_sym}
	Let $\Pi$ be a multiplier and $(\Psi,M)$ any factorization with $\Psi$ stable. Assume $\Delta_2$ is a causal bounded operator such that
	\begin{equation}\label{eq:symIQC}
	\int_{-\infty}^{\infty}
	\begin{bmatrix}\widehat{v}(j\omega)\\ \widehat{w}(j\omega)\end{bmatrix}^*
	\Pi(j\omega)
	\begin{bmatrix}\widehat{v}(j\omega)\\ \widehat{w}(j\omega)\end{bmatrix}
	d\omega\leq -\epsilon (\|v\|^2 + \|w\|^2), 
	\end{equation}	
        for any $w\in\mathcal{L}_2[0,\infty)$ and $v=\Delta_2 w$.  Then for all
        $T\geq0$, for all $w\in\mathcal{L}_2[0,\infty)$, and $v=\Delta_2 w$, the
        signal defined by $z=\Psi\begin{bmatrix}v\\ w\end{bmatrix}$
        satisfies
	\begin{equation}
	\int_{0}^{T}
	z(t)^\top 
	M z(t)
	dt\leq -\epsilon \|z\|_T  -\underbar{J}(x(T)),
	\end{equation}
	where $x(T)$ denotes the state of the system $\Psi$ at the instant $T$ when driven by the inputs $(v,w)$ with null initial conditions. 
\end{lem1}
\textbf{Proof:} See Appendix.\qed

\begin{thm1}\label{thm:1}
Given a positive-negative multiplier $\Pi\in\mathbf{RL}_\infty$, the factorization $(\Psi,M)$ is doubly-hard if $\Psi,\Psi^{-1}\in\mathbf{RH}_\infty$. 
\end{thm1}
\textbf{Proof: } If $\Psi\in\mathbf{RH}_\infty$, Lemma~\ref{lem_ori} and Lemma~\ref{lem_sym} hold. Moreover, if the multiplier $\Pi$ is positive-negative and $\Psi^{-1}\in\mathbf{RH}_\infty$ then $\bar{J}(x)=\underbar{J}(x)=0$ for all $x\in\mathds{R}^n$ (see Lemma 5 in~\citep{Seiler:15}). As a result the factorization $(\Psi,M)$ is doubly-hard.	
\qed

Therefore all $J$-spectral factorizations are doubly-hard factorizations.

\section{Triangular factorization \emph{vs} $J$-spectral factorization }

This section provides a simple example highlighting the distinction
between triangular and $J$-spectral factorizations.  Consider a simple
feedback interconnection of the static system $G=\frac{1}{2}$ and an
operator $\Delta$. Define an positive-negative multiplier 
$\Pi\in\mathbf{RL}_\infty^{2 \times 2}$ by
\begin{align}
\Pi(s) = \begin{bmatrix} 3 & \frac{-s+2}{s+1} \\
\frac{-s-2}{s-1} & \frac{-s^2+4}{s^2-1} \end{bmatrix}
\end{align}
Assume the interconnection of $G$ and $\tau\Delta$ is well-posed for
all $\tau \in [0,1]$.  Also assume that $\tau \Delta$ satisfies the
IQC defined by $\Pi$ for all $\tau \in [0,1]$.  It can be verified
that $\bsmtx G \\ 1 \esmtx^\sim \Pi \bsmtx G \\ 1 \esmtx = -1.25<0$,
i.e. $G$ satisfies the IQC constraint with $\Pi$. Thus the frequency
domain IQC conditions in Theorem~\ref{thm:IQC} are satisfied and the
feedback interconnection is stable.

\begin{figure}[!h]
	\centering
	\includegraphics[width=3.2in]{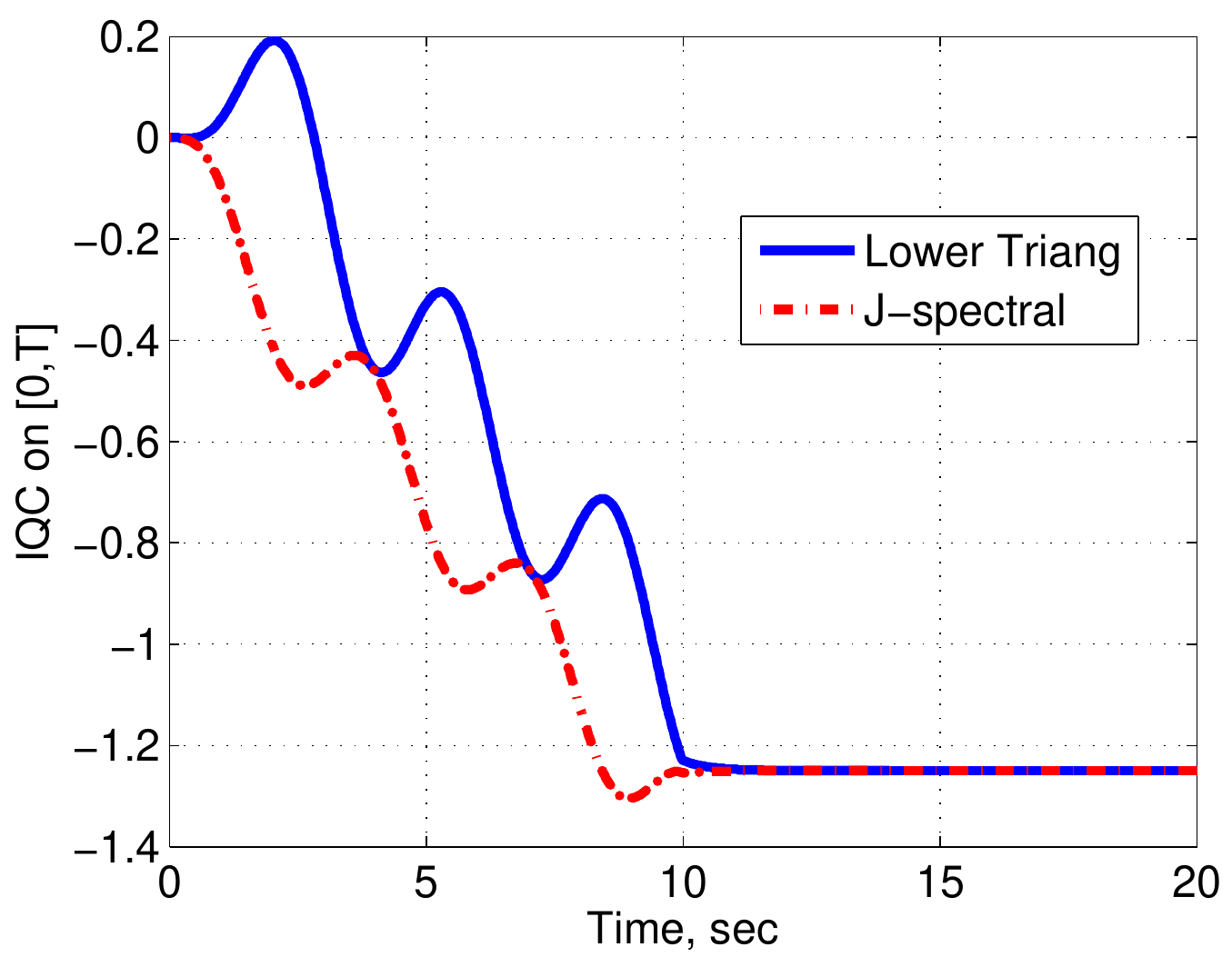}
	\caption[IQC Plot]{
		$\int_0^T \left( \Psi \bsmtx G \\ 1 \esmtx u \right)^T M \left(
		\Psi \bsmtx G \\ 1 \esmtx u \right) dt$ versus time $T$.}
	\label{fig:IQCplot}
\end{figure}

As noted above, the factorization of $\Pi$ is not unique.  Here we
construct two different factorizations.  First, a stable triangular
factorization $(\Psi,M)$ of $\Pi$ is given by:
\begin{equation}
M=\bsmtx 1 & 0 \\ 0 & -1 \esmtx
\mbox{ and } \Psi = \bsmtx 2 & 0 \\ 1 & \frac{s-2}{s+1} \esmtx.
\end{equation}
Note that $\Psi$ is stable but the (2,2) entry of $\Psi$ is
non-minimum phase.  The multiplier satisfies the positive-negative
conditions and hence it also has a $J$-spectral factorization
$(\tilde \Psi, \tilde J)$:
\begin{equation}
\tilde J=\bsmtx 1 & 0 \\ 0 & -1 \esmtx
\mbox{ and }
\tilde \Psi = \bsmtx -1.751 & \frac{0.4133 s-1.508}{s+1} \\ -0.2554 &
\frac{-1.082s-2.505}{s+1} \esmtx.
\end{equation}
Note that for this factorization $\tilde \Psi$ and $\tilde
\Psi^{-1}$ are both stable.  Figure~\ref{fig:IQCplot} shows the IQC
evaluated on $[0,T]$ versus the finite horizon time $T$ for the input
signal $u(t)=0.458\sin(t)$ for $t \in [0,10]$ and $u(t)=0$
otherwise.  The coefficient $0.458$ is selected to normalize the
signal $\|u\|_2 = 1$.  As $t\rightarrow \infty$, both IQCs converge to
$-1.25$.  This value is consistent with the constraint $\bsmtx G \\ 1
\esmtx^\sim \Pi \bsmtx G \\ 1 \esmtx = -1.25 < 0$.  Thus both
factorizations satisfy the time-domain constraint as $t \rightarrow
\infty$.  However, the lower triangular factorization goes positive
on the approximate interval $[0,2.8]$.  Thus the lower triangular
factorization can violate the constraint over finite horizons.  On the
other hand, the $J$-spectral factorization remains negative and hence
satisfies the constraint over all finite horizons.

It can be shown that lower triangular factorizations 
have a $(2,2)$ entry that is non-minimum phase in general. Specifically,
if $\Psi$ is lower triangular and $\Pi = \Psi^\sim J \Psi$ then
the entries of $\Psi$ satisfy:
\begin{align*}
\Pi_{11} & = \Psi_{11}^\sim \Psi_{11} - \Psi_{21}^\sim \Psi_{21} \\
\Pi_{12} & = -\Psi_{21}^\sim \Psi_{22} \\
\Pi_{22} & = -\Psi_{22}^\sim \Psi_{22}
\end{align*}
These conditions imply that if $\Pi_{12}$ has poles in the left half
plane then $\Psi_{22}$ must be non-minimum phase. Specifically, if
$\Pi$ is a positive-negative multiplier then there exists $\epsilon>0$
such that $-\Pi_{22}(j\omega)\ge \epsilon I$ $\forall \omega \in R$.
Hence it can be factorized as $-\Pi_{22} = H^\sim H$ where
$H \in RH$ and $H^{-1}$ is anti-stable. In other words $H$
is stable and anti-minimum phase.  This factorization can be
constructed from the normal stable, minimum phase spectral
factorization \citep{youla61}.  Next, let $\{p_i\}_{i=1}^N$ denote the
poles of $\Pi_{12}$ in the left half plane.  Define $\Psi_{21}$ and
$\Psi_{22}$ as
\begin{align}
\Psi_{22}(s) & := H(s) \left( \prod_{i=1}^N \frac{s+\bar{p}_i}{s-p_i} 
\cdot I_m \right) \\
\Psi_{21}(s) & := \left( -\Pi_{12}(s) \Psi_{22}^{-1}(s) \right)^\sim
\end{align}
By construction, $\Psi_{22}$ is stable and anti-minimum phase.  The
inclusion of the Blaschke products\footnote{See \citep{partington2004} for a definition.} in the definition of $\Psi_{22}$
does not impact the value of $\Psi_{22}^\sim \Psi_{22}$ on the
imaginary axis.  Thus $\Pi_{22} = \Psi_{22}^\sim \Psi_{22}$ on the
imaginary axis by construction of $H$.  This choice of $\Psi_{22}$ is
required to ensure that $\Pi_{12} \Psi_{22}^{-1}$ is anti-stable and
hence $\Psi_{21}$ is stable.  Moreover, $\Psi_{21}^\sim \Psi_{22}=-\Pi_{12}$.
A stable, stably invertible $\Psi_{11}$ can then be constructed from
a spectral factorization of $\Pi_{11}+\Psi_{21}^\sim \Psi_{21}$.
In this construction, any LHP poles of $\Pi_{12}$ appear as RHP
zeros in $\Psi_{22}$.

\section{Comparison between IQC and graph separation results}

\subsection{Stability results}

In this section we develop two stability results for the feedback interconnection of two nonlinear systems. One of these results will be obtained using graph separation methods. For completeness, we state an IQC version of Corollary 5.1 in~\citep{Teel:96} as follows:

\begin{thm1}[\cite{Teel:96}]\label{thm:graph}
	Let $\Delta_1$ and $\Delta_2$ be two causal and bounded systems. Let $\Psi$ be a stable linear system. Assume that:
	\begin{enumerate}
		\item the feedback interconnection of $G$ and $\Delta$ is well-posed;
		\item the time-domain IQC 
		\begin{equation}
		\int_{0}^{T}\left(\Psi\begin{bmatrix}u\\ \Delta_1 u\end{bmatrix}\right)^\top  M \left(\Psi\begin{bmatrix}u\\ \Delta_1  u\end{bmatrix} \right) dt\geq 0,
		\end{equation}
		is satisfied for any $T>0$ and $u\in\mathcal{L}_{2e}[0,\infty)$;
		\item the time-domain inverse-graph IQC  
		\begin{equation}
		\int_{0}^{T}\left(\Psi\begin{bmatrix}\Delta_2u\\ u\end{bmatrix}\right)^\top 
		M \left(\Psi\begin{bmatrix}\Delta_2u\\ u\end{bmatrix}\right) dt
		< -\epsilon \left\|\begin{bmatrix}\Delta_2u\\ u\end{bmatrix}\right\|_T^2,
		\end{equation}	
		is satisfied for any $T>0$ and $u\in\mathcal{L}_{2e}[0,\infty)$.
	\end{enumerate}
	Then the feedback interconnection between $\Delta_1$ and $\Delta_2$ is $\mathcal{L}_2$-stable.
\end{thm1}

In the spirit of~\cite{Jonsson:11}, we can establish the following corollary for the interconnection of two nonlinear systems:
\begin{cor1}[Corollary of Theorem~\ref{thm:IQC}]\label{Cor:IQC}
	  Let $\Delta_1:\mathcal{L}^l_{2e}[0,\infty)\rightarrow\mathcal{L}^m_{2e}[0,\infty)$ and
	  $\Delta_2:\mathcal{L}^m_{2e}[0,\infty)\rightarrow\mathcal{L}^l_{2e}[0,\infty)$
	  be bounded causal operators, and let
	  $\Pi\in\mathbf{RL}_\infty^{(m+l)\times (m+l)}$. Assume that:
	  \begin{enumerate}
	  	\item[(I)] for every $\tau\in[0,1]$, the feedback interconnection of $\tau\Delta_1$ and $\tau\Delta_2$ is well-posed;
	  	\item[(II)] for every $\tau\in[0,1]$, $\tau\Delta_1$ satisfies the IQC defined by $\Pi$;
	  	\item[(III)] for every $\tau\in[0,1]$, $\tau\Delta_2$ strictly satisfies the inverse-graph IQC defined by $\Pi$.  		
	  \end{enumerate}
	  Then, the feedback interconnection of $\Delta_1$ and $\Delta_2$ is stable.
\end{cor1}
\textbf{Proof:} The result follows from the application of the IQC theorem using 
\begin{equation}
\Delta=\begin{bmatrix}
\Delta_1 & 0\\
0        & \Delta_2
\end{bmatrix} \text{ and } G=\begin{bmatrix}
0 & I\\
I & 0
\end{bmatrix}, 
\end{equation}
and the following augmented multiplier:
\begin{equation}
\Pi_a=\begin{bmatrix} \Pi_{11} & 0 & \Pi_{12} & 0\\
0  &  -\Pi_{22}-\epsilon I & 0 & -\Pi_{12}^*\\
\Pi_{12}^* & 0 & \Pi_{22} & 0\\
0  &  -\Pi_{12} & 0 & -\Pi_{11}^*-\epsilon I\\
\end{bmatrix}
\end{equation}
Some straightforward algebra is required to show that the conditions in Theorem~\ref{thm:IQC} are satisfied.\qed

Using Theorem~\ref{thm:1}, then it is possible to remove the homotopy condition in the above result if the matrix $\Pi$ is positive-negative. Formally we can state the following result:
\begin{cor1}[Corollary of Theorem~\ref{thm:graph}]~\label{Cor:GS}
	Let $\Delta_1:\mathcal{L}^l_{2e}[0,\infty)\rightarrow\mathcal{L}^m_{2e}[0,\infty)$ and
	$\Delta_2:\mathcal{L}^m_{2e}[0,\infty)\rightarrow\mathcal{L}^l_{2e}[0,\infty)$
	be bounded causal operators, and let
	$\Pi\in\mathbf{RL}_\infty^{(m+l)\times (m+l)}$. Assume that:
	\begin{enumerate}
		\item[(i)] the feedback interconnection of $\Delta_1$ and $\Delta_2$ is well-posed;
		\item[(ii)] $\Delta_1$ satisfies the IQC defined by $\Pi$;
		\item[(iii)] $\Delta_2$ strictly satisfies the inverse-graph IQC defined by $\Pi$;
		\item[(iv)] $\Pi$ is a positive-negative multiplier.
	\end{enumerate}
	Then, the feedback interconnection of $\Delta_1$ and $\Delta_2$ is stable.
\end{cor1}  
\textbf{Proof:} If $\Pi$ is a positive-negative multiplier, then there exists a factorization $(\Psi,M)$ such that $\Psi$ and $\Psi^{-1}$ are both stable~\citep{Seiler:15}. Therefore the factorization $(\Psi,M)$ is doubly-hard as it satisfies the conditions in Theorem~\ref{thm:1}. The frequency-domain conditions (ii) and (iii) can be transformed into truncated time-domain conditions by using the factorization $(\Psi,M)$. As a result, Theorem~\ref{thm:graph} can be used to establish the stability of feedback interconnection between $\Delta_1$ and $\Delta_2$.
\qed

\begin{rem1}
	It would not be possible to prove Corollary 2 by using triangular factorizations as it fails to guarantee that condition (iii) is equivalent to the truncated time-domain condition~(28).
\end{rem1}

\subsection{Discussion}

A na\"ive comparison of the results would suggest that condition (iv) in Corollary 2  an extra condition over the conditions of Corollary 1. It is well known that the homotopy condition in (II) is satisfied if $\Pi_{11}$ is positive. Similarly, the homotopy condition in (III) is satisfied if $\Pi_{22}$ is negative. Hence one can think of a superiority of Corollary 1 over Corollary 2.

However, if $\Delta_1$ and $\Delta_2$ are both nonlinear, the IQC theorem requires homotopy conditions for both systems. If condition (II) holds, the requirement of the condition to be true when $\tau=0$ implies $\Pi_{11}(j\omega)\geq 0$ for all $\omega\in\mathds{R}$. Similarly, if condition (III) holds, the same argument when $\tau=0$ implies $\Pi_{22}(j\omega)\leq-\epsilon I$ for some $\epsilon>0$.

A perturbation argument as in~\citep{Carrasco12a,Seiler:15} in conjunction with a substitution argument~\citep{Carrasco:13} is required here; although $\Pi_{11}(j\omega)\geq 0$ does not guarantee the existence of a factorization, the following Lemma ensures the existence of a new $\bar{\Pi}$ with $\bar{\Pi}_{11}(j\omega)\geq \delta I$ for some $\delta>0$, hence $\bar{\Pi}$ can be factorised:
\begin{lem1}\label{lem:pert}   Let $G\in\mathbf{RH}^{m\times l}_\infty$, let
	$\Delta:\mathcal{L}^m_{2e}[0,\infty)\rightarrow\mathcal{L}^l_{2e}[0,\infty)$
	be a bounded causal operator. If conditions (2) and (3) in Theorem~\ref{thm:IQC} are  satisfied for some $\Pi$, then there exists some $\delta>0$ such that conditions (2) and (3) are satisfied for 
	$$	\bar{\Pi}=\begin{bmatrix}
	\Pi_{11}+\delta I_{m} & \Pi_{12}\\
	\Pi_{21} & \Pi_{22}
	\end{bmatrix}.$$
\end{lem1}
\textbf{Proof:} See Appendix.\qed
\begin{rem1}
	The counterpart result for Corollary~\ref{Cor:IQC} is trivially obtained as the only required condition that is the boundedness of $\Delta_2$.
\end{rem1}

As a result, we can consider without loss of generality that Corollary~\ref{Cor:IQC} can only be satisfied if $\Pi$ is positive-negative. In conclusion, the IQC theorem may only provide better results over the graph separation theory when (a) $\Delta_2$ is linear and (b) $\Pi_{22}$ is non-negative. Otherwise, graph separation and IQC theories lead to the same stability result for rational multipliers. 

\section{Conclusion}

The aim of this paper is to complete the classification of IQC-factorizations. It concludes previous work presented in~\citep{Seiler:15,Carrasco:15}, establishing a novel connection between IQC and graph separation theories. Here we propose the term doubly-hard factorizations, where both frequency conditions can be transformed into truncated time-domain conditions. We show that the standard triangular factorization is hard factorization but fails to be a doubly-hard. Then it cannot be used to establish an equivalence between the IQC theorem and separation results in the truncated time-domain. We have shown that $(\Psi,M)$ is a doubly-hard factorization if $\Psi$ and $\Psi^{-1}$ are both stable. 

The new results allow us to compare both theories for the feedback interconnection two nonlinear systems. As a result we conclude that the IQC theorem for two nonlinear system does not provide any significant advantage over its counterpart result derived using graph separation tools. However, the IQC theorem may provide some advantages when one of the system is linear and the term $\Pi_{22}$ is non-negative.

\section*{Acknowledgement}

The first author acknowledges William Heath for fruitful discussions and comments. 

\bibliographystyle{plainnat}        
\bibliography{autosam}           
\appendix
\section{Proof of Lemma~\ref{lem:pert}}
If condition (2) is satisfied for $\Pi$, then it is trivial that it is also satisfied for $\bar{\Pi}$ since
\begin{equation}
\begin{bmatrix}\widehat{u}(j\omega)\\ \widehat{\Delta u}(j\omega)\end{bmatrix}^*
\begin{bmatrix}
\delta I_{m} & 0\\
0 & 0
\end{bmatrix}
\begin{bmatrix}\widehat{u}(j\omega)\\ \widehat{\Delta u}(j\omega)\end{bmatrix}=\delta|\widehat{u}(j\omega)|^2\geq 0
\end{equation}
for all $\omega\in\mathds{R}$.
If condition (3) is satisfied for $\Pi$, then there exists $\epsilon>0$ such that
\begin{equation}
\begin{bmatrix}
G(j\omega)\\I 
\end{bmatrix}^*\Pi(j\omega)\begin{bmatrix}
G(j\omega)\\I 
\end{bmatrix} \leq -\epsilon I\qquad \forall\omega\in\mathds{R}.
\end{equation}
Moreover, for any $\delta>0$, it follows
\begin{equation}
\begin{bmatrix}
G(j\omega)\\I 
\end{bmatrix}^*
\begin{bmatrix}
\delta I_{m} & 0\\
0 & 0
\end{bmatrix}
\begin{bmatrix}
G(j\omega)\\I 
\end{bmatrix}=\delta(G(j\omega)^*G(j\omega))\leq\delta \|G\|^2_\infty I,
\end{equation}
for all $\omega$. As a result, taking $\delta=\frac{\epsilon}{2\|G\|^2_\infty}$, 
\begin{equation}
\begin{bmatrix}
G(j\omega)\\I 
\end{bmatrix}^*\bar{\Pi}(j\omega)\begin{bmatrix}
G(j\omega)\\I 
\end{bmatrix} \leq -(\epsilon-\frac{\epsilon}{2}) I=-\frac{\epsilon}{2} I\qquad \forall\omega\in\mathds{R},
\end{equation}
\section{Proof of Lemma~\ref{lem_sym}}

For any $T\geq 0$, the frequency-domain inequality~\eqref{eq:symIQC}
can be converted into time-domain (by Parserval's theorem) and
re-arranged as
\begin{equation}
\int_{0}^{T}z(t)^\top M z(t) dt\leq -\epsilon\int_{0}^{\infty}z(t)^\top z(t)dt-\int_{T}^{\infty}z(t)^\top M z(t) dt
\end{equation}
Note that $-\epsilon\int_{T}^{\infty}z(t)^\top z(t)dt \le 0$ and hence
the following bound is also valid:
\begin{equation}
\label{eq:appbound}
\int_{0}^{T}z(t)^\top M z(t) dt\leq -\epsilon\int_{0}^{T}z(t)^\top z(t)dt-\int_{T}^{\infty}z(t)^\top M z(t) dt
\end{equation}
Next let $\tilde{w}\in\mathcal{L}_2[0,\infty)$ be any signal satisfying
$\tilde{w}_T=w_T$. Define $\tilde{v}=\Delta_2 \tilde{w}$ and let
$\tilde{z}=\Psi\left[\begin{smallmatrix}\tilde{v}\\
    \tilde{w}\end{smallmatrix} \right]$ be the response of $\Psi$ with
null initial condition. By causality of $\Delta_2$ and $\Psi$,
$w_T=\tilde{w}_T$ implies $v_T=\tilde{v}_T$ and
$z_T=\tilde{z}_T$. Hence for all $\tilde{w}$, it holds

$$\int_{0}^{T}z(t)^\top M z(t) dt=\int_{0}^{T}\tilde{z}(t)^\top M \tilde{z}(t) dt.$$

Moreover, the IQC holds for any input/output pairs of $\Delta_2$. In
particular, Equation~\ref{eq:appbound} holds with $z$ replaced by
$\tilde{z}$.  As a result, any $\tilde{w}\in\mathcal{L}_2$ satisfying
$\tilde{w}_T = w_T$ can be used to upper bound the integral
$\int_{0}^{\infty}z(t)^\top M z(t) dt$ obtained with $w$:
\begin{equation}
\int_{0}^{T}z(t)^\top M z(t) dt
   \leq -\epsilon\int_{0}^{T} z(t)^\top z(t)dt
        -\int_{T}^{\infty} \tilde{z}(t)^\top M \tilde{z}(t) dt
\end{equation}
Minimizing over all feasible $\tilde{w}$ yields the upper bound
\begin{multline}
\int_{0}^{T}z(t)^\top M z(t) dt\leq 
-\epsilon\int_{0}^{T}z(t)^\top z(t)dt\\
+\inf_{\tilde{w}\in\mathcal{L}_2, \tilde{w}_T=w_T}\left(-\int_{T}^{\infty}\tilde{z}(t)^\top M \tilde{z}(t) dt\right),
\end{multline}
The suitable set of signals $\tilde{w}$ can be rewritten as 
$$\tilde{w}(t)=\begin{cases}
w(t) & \text{ if } t\leq T\\
w_f(t) & \text{ if } t>T 
\end{cases}
$$
for any $w_f\in\mathcal{L}_2[T,\infty)$. We can rewrite the minimisation as
\begin{multline}
\label{eq:appbound2}
\int_{0}^{T}z(t)^\top M z(t) dt\leq -\epsilon\int_{0}^{T}z(t)^\top z(t)dt+\\\inf_{w_f\in\mathcal{L}_2[T,\infty)}\left(-\int_{T}^{\infty}\tilde{z}(t)^\top M \tilde{z}(t) dt\right),
\end{multline}
such that $\tilde{v}=\Delta_2 \tilde{w}$ and
$\tilde{z}=\Psi\begin{bmatrix}\tilde{v}\\ \tilde{w}\end{bmatrix}$.
The dependence on $\Delta_2$ can be removed following similar
arguments to those given in ~\citep{Seiler:15}. Partition
$\tilde{v}=\Delta_2 \tilde{w}$ as:
\begin{equation}
  \tilde{v}(t)=\begin{cases}
    \Delta_2 w(t) & \text{ if } t\leq T\\
    v_f(t) & \text{ if } t>T 
\end{cases}
\end{equation}
The bound in Equation~\ref{eq:appbound2} only involves $\tilde{z}$
defined on $[T,\infty)$.   This signal can be computed from the
state of $\Psi$ at time $T$, i.e. $x_T$, as well as the signals
$w_f$ and $v_f$.  Note that $x(T)=x_T$ is the same for any feasible choice
of $\tilde{w}$ because $\tilde{w}_T=w_T$ and $\tilde{v}_T=v_T$.  
The dependence on $\Delta_2$ is removed, with some conservatism, 
by simply maximizing over all possible future signals $v_f$ defined
on $[T,\infty)$ instead of using
$\tilde{v}=\Delta_2 w$.  In other words,
\begin{multline}
\int_{0}^{T}z(t)^\top M z(t) dt\leq -\epsilon\int_{0}^{T}z(t)^\top z(t)dt+\\\inf_{w_f\in\mathcal{L}_2[T,\infty)}\sup_{v_f\in\mathcal{L}_2[T,\infty)} \left(-\int_{T}^{\infty}\tilde{z}(t)^\top M \tilde{z}(t) dt\right),
\end{multline}
This is subject to constraint $x(T)=x_T$.  This can be rewritten
using the cost function $J$ as:
\begin{multline}
\int_{0}^{T}z(t)^\top M z(t) dt\leq 
-\epsilon\int_{0}^{T}z(t)^\top z(t)dt-\underbar{J}(x_T),
\end{multline}
\end{document}